\def\year{2022}\relax
\documentclass[letterpaper]{article} 
\usepackage{aaai22}  
\usepackage{times}  
\usepackage{helvet}  
\usepackage{courier}  
\usepackage[hyphens]{url}  
\usepackage{graphicx} 
\usepackage{natbib}  
\usepackage{caption} 
\DeclareCaptionStyle{ruled}{labelfont=normalfont,labelsep=colon,strut=off} 
\usepackage{algorithm}
\usepackage{algorithmic}

\usepackage{newfloat}
\usepackage{listings}
\floatstyle{ruled}
\newfloat{listing}{tb}{lst}{}
\floatname{listing}{Listing}

\usepackage{multirow}
\usepackage{subfigure}

\title{MonaCoBERT: Monotonic attention based ConvBERT for Knowledge Tracing}
\author{
    Unggi Lee\textsuperscript{\rm 1},
    Yonghyun Park\textsuperscript{\rm 2},
    Yujin Kim\textsuperscript{\rm 3}, \\
    Seongyune Choi\textsuperscript{\rm 1},
    Hyeoncheol Kim\textsuperscript{\rm 1}\footnote{Corresponding author.}
}
\affiliations{
    \textsuperscript{\rm 1}Department of Computer Science and Engineering, Korea University, Seoul, Republic of Korea\\
    \textsuperscript{\rm 2}Department of Physics Education, Seoul National University, Seoul, Republic of Korea\\
    \textsuperscript{\rm 3}Department of Science Education, Ewha Womans University, Seoul, Republic of Korea\\

    codingchild@korea.ac.kr,
    enkeejunior1@snu.ac.kr,
    hello.yujink@gmail.com, \\
    \{csyun213, harrykim\}@korea.ac.kr
}

\usepackage{bibentry}

\begin{document}

\maketitle 

\begin{abstract}

Knowledge tracing (KT) is a field of study that predicts the future performance of students based on prior performance datasets collected from educational applications such as intelligent tutoring systems, learning management systems, and online courses. Some previous studies on KT have concentrated only on the interpretability of the model, whereas others have focused on enhancing the performance. Models that consider both interpretability and the performance improvement have been insufficient. Moreover, models that focus on performance improvements have not shown an overwhelming performance compared with existing models. In this study, we propose MonaCoBERT, which achieves the best performance on most benchmark datasets and has significant interpretability. MonaCoBERT uses a BERT-based architecture with monotonic convolutional multihead attention, which reflects forgetting behavior of the students and increases the representation power of the model. We can also increase the performance and interpretability using a classical test-theory-based (CTT-based) embedding strategy that considers the difficulty of the question. To determine why MonaCoBERT achieved the best performance and interpret the results quantitatively, we conducted ablation studies and additional analyses using Grad-CAM, t-SNE, and various visualization techniques. The analysis results demonstrate that both attention components complement one another and that CTT-based embedding represents information on both global and local difficulties. We also demonstrate that our model represents the relationship between concepts.

\end{abstract}

\section{Introduction}

The outbreak of COVID-19 has accelerated the digital transformation in the field of education, and the number of students who use online learning platforms has increased. Most online learning platforms collect student data such as interaction logs, correctness, and learning history, thereby providing a chance to develop a better adaptive learning system for students. 

Knowledge tracing (KT) is a research area that predicts the future performance of students based on prior performance datasets collected from educational applications such as intelligent tutoring systems (ITS), learning management systems (LMS), and online courses. KT models can be broadly classified into two categories: those focusing on interpretability and those focusing on increasing the performance. Models focused on interpretability, such as BKT \cite{corbett1994knowledge} and PFA \cite{pavlik2009performance}, mainly use Markov chain techniques and logistic functions. These models are generally simple to interpret. However, they suffer from a relatively low performance \cite{piech2015dkt}. Models focusing on a performance improvement, such as DKT, DKVMN \cite{zhang2017dynamic}, SAKT \cite{pandey2019self}, and CL4KT \cite{lee2022contrastive}, perform significantly better than traditional statistical approaches. However, this is difficult to interpret because of the nature of deep learning. Although AKT \cite{ghosh2020context} considers both the model performance and interpretability, it has not shown an overwhelming performance in comparison to existing models. 

Moreover, despite the application of self-attention and transformers \cite{vaswani2017attention} in recent KT models, they have been unable to improve the attention architecture compared to changes in other parts of the models. In natural language processing (NLP), Bigbird \cite{zaheer2020big}, Longformer \cite{beltagy2020longformer}, and ConvBERT \cite{jiang2020convbert}, which maintain the BERT architecture and change the attention architectures, have succeeded in terms of both performance and efficiency.

In this article, MonaCoBERT, which achieves both a high performance and interpretability, is proposed. MonaCoBERT uses BERT-based and monotonic convolutional attention architectures. We also suggest a classical test theory (CTT) based embedding strategy that considers the question difficulty. Using CTT-based embedding, our model achieves a increase in performance and interpretability. As a result, MonaCoBERT achieved state-of-the-art results on most benchmark datasets in term of both the AUC and RMSE. Moreover, we also conducted an ablation study on all parts of the models and an additional analysis using Grad-CAM, t-SNE, and various visualization techniques. The analysis results demonstrate that both attention components complement one another and that CTT-based embedding represents information on both global and local difficulties. We also demonstrate that our model represents the relationship between concepts.

\vspace{\bigskipamount}

\begin{figure*}[hbt!]
\centering
\includegraphics[width=1.8\columnwidth]{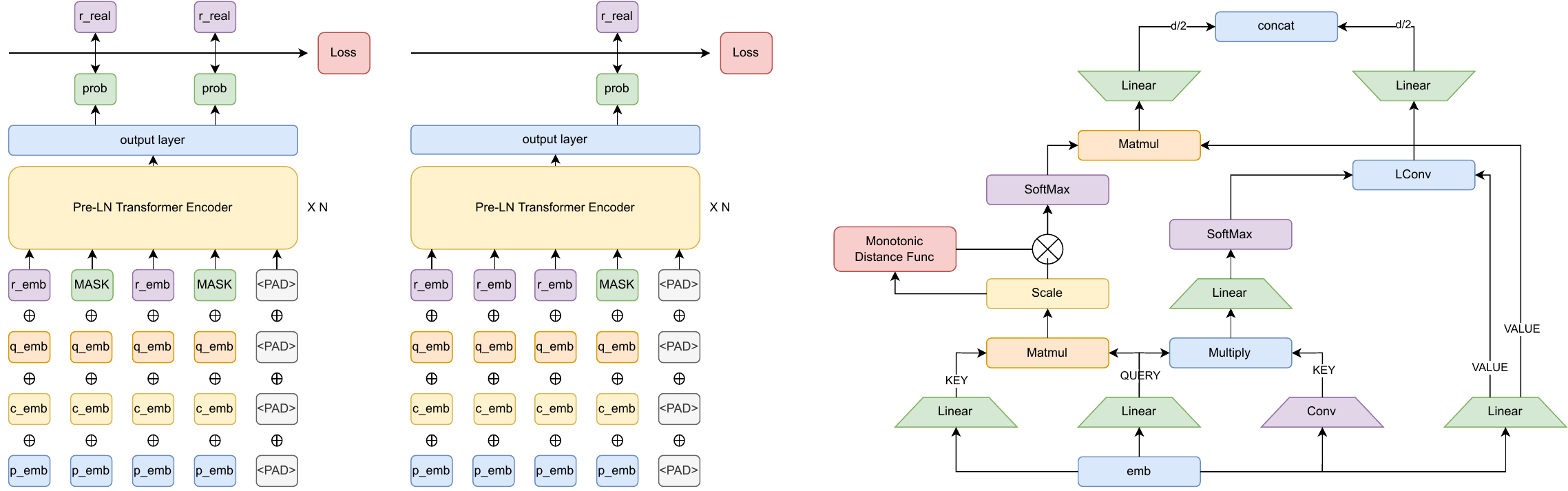}
\caption{Architectures of MonaCoBERT and monotonic convolutional multi-head attention. The \textit{left} side shows different strategies in training and testing sessions. The \textit{right} side shows the architecture of monotonic convolutional multi-head attention, combined with monotonic attention and ConvBERT attention.}
\label{monoconvattn}
\end{figure*}

\section{Related Work}

\subsection{Knowledge Tracing}
Knowledge tracing (KT) is a research area of predicting the knowledge states of the students using their interaction data. Since the first introduction of DKT \cite{piech2015dkt}, significant research in this area using deep neural networks has been conducted. Researchers have recently focused on self-attention architectures. SAKT \cite{pandey2019self}, SAINT+ \cite{shin2021saint+}, which uses self-attention, achieves a better performance than previous models. Moreover, AKT \cite{ghosh2020context} was presented with self-attention, and a new architecture for retrieving latent knowledge representations was suggested. For AKT, a new embedding method that considers the educational perspective has also been suggested. CL4KT \cite{lee2022contrastive} also uses a self-attention and contrastive learning framework, and has achieved the best performance in KT.

\subsection{BERT and Its Application}
BERT \cite{devlin2018bert} has been referred to as a successful application of Transformer. It mainly uses self-attention, and the masked language model (MLM) method, which can train bidirectionally in NLP, has been suggested. Some variations of BERT, such as Bigbird \cite{zaheer2020big}, Longformer \cite{beltagy2020longformer}, and ConvBERT, \cite{jiang2020convbert} have recently been designed with effective attention mechanisms applied while using the original architecture of BERT. These approaches have achieved an outstanding performance and high efficiency. Other studies have also attempted to use BERT architectures. As a recommendation system, BERT4Rec \cite{sun2019bert4rec} uses the BERT architecture to enhance the recommendation power. However, for KT, although BEKT \cite{tianabekt} and BiDKT \cite{tan2022bidkt} attempt to use the BERT architecture, they cannot achieve a higher performance than other KT models. In this study, we explored why BERT does not perform better than the other models and showed that the BERT architecture is still valuable for KT. We found that changing the attention architecture and an embedding strategy are vital to optimizing BERT for the KT area.

\section{Method}

\subsection{Problem Statement}
Knowledge tracing aims at predicting the probabilities of students being correct through the use of a sequence of interaction data gathered by an LMS or ITS. Student interactions can be expressed as $x_1, ..., x_t,$, and the $t$-th interaction can be denoted as $x_t = (q_t, a_t)$. Here, $q_t$ is the t-$th$ question and $a_t$ is $t$-th correctness of the student's response, where $a_t \in \{0, 1\}$, in which 0 indicates an incorrect response and 1 is a correct answer. However, some datasets contain concept data $c_t$, and thus we can also express  $x_t = (q_t, c_t, a_t)$.

\subsection{Proposed Model Architecture}

\subsubsection{BERT based Architecture for Knowledge Tracing}
To create our model baseline, we mainly referenced BERT \cite{devlin2018bert}, BiDKT \cite{tan2022bidkt}, BEKT \cite{tianabekt}, and BERT4Rec \cite{sun2019bert4rec}. To optimize our research into KT, we changed some of the BERT architecture. First, we used a pre-layer normalization (pre-LN) Transformer in our model. Previous research \cite{liu2020understanding} has suggested that Transformer is difficult to train without a training strategy, such as a warm-up start. By contrast, the pre-LN Transformer can be trained without a warm-up start and converges much faster than the original Transformer \cite{xiong2020layer}. Second, we used a different strategy for the training and testing processes. During the training process, the proposed model predicted the masking position. The masking ratio used in the training process was the same as with the original BERT, which used 15\% embedding, 80\% of which was actual masking, 10\% was a reversal, and 10\% did not change. During the testing process, masking was applied to the last position of each sequence. Referring to the previous BERT-based studies on KT \cite{tan2022bidkt} or recommendation systems \cite{sun2019bert4rec}, the model predicts the correctness of the students using their previous history of interaction sequences. Figure \ref{monoconvattn}-\textit{Left} describes the different training and testing strategies of our model.

\subsubsection{Embedding Strategy}
Most KT models use concepts, questions, and correctness as the input vectors for training. Previous studies have explored new input features. For example, AKT created Rasch embedding vectors by using concepts, items, and responses \cite{ghosh2020context}. However, an item response theory (IRT), such as Rasch, can be applied to the dataset collected from tests or examinations because IRT assumes that the ability of a student does not change during the trial. In KT, the states of student knowledge change during learning \cite{yeung2019deep}. Therefore, we used the classical test theory (CTT) for handling the difficulty features.

We extracted the correctness of each question from the training set and made the questions difficulty. If the question in the validation or test set were not contained in training set, we replaced that question difficulty as a arbitrarily number like 75. Subsequently, we added the difficulty to the embedding blocks. In a previous study, BEKT \cite{tianabekt} used five difficulty ranges in its embedding blocks. Nevertheless, we used a difficulty range of 100.  Similar to BERT embedding layers, MonaCoBERT uses element-wise embedding vectors $E_{input}$, learnable positional embedding $E_{pos}$, concept embedding $E_c$, item embedding $E_q$, correctness embedding $E_a$, and CTT embedding $E_{ctt}$, where $E_{input} \in R^{m \times h},\, E_{pos} \in R^{m \times h},\, E_c \in R^{m \times h},\, E_q \in R^{m \times h},\, E_a \in R^{m \times h}\, and\, E_{ctt} \in R^{m \times h}$. Embedding layers $E_{input}$ are formulated as follows: 

\begin{equation}
    E_{input} = E_{pos} + E_c  +  E_q + E_a + E_{ctt}
\end{equation}

\subsubsection{Pre-LN Transformer-based Encoder Architecture}
The encoder blocks used the pre-LN Transformer architecture \cite{xiong2020layer}. In this study, 12 encoder layers were used. First, the embedding vectors $E_{input}$ are normalized through the pre-LN $LN_{pre}$

\begin{equation}
    z = LN_{pre}(E_{input})
\end{equation}

Second, the normalized value $z$ was changed to the query, key, and value of monotonic convolutional multihead attention. The results were passed through dropout layer $D$ and added to the embedding vectors as a residual connection.

\begin{equation}
    a = x + D(MonoConvMulAttn(z, z, z))
\end{equation}

Third, the results were normalized and passed through fully connected layers. The activation function was LeakyReLU. The results were also normalized through the dropout layer $D$. The second result was added as a residual connection.

\begin{equation}
    l = a + D(fc(LN(a)))
\end{equation}

The fully connected layers are formulated as follows. 

\begin{equation}
    fc = W_{fc2}(LeakyReLU(W_{fc1})),
\end{equation}

where $W_{fc1} \in R^{h \times (h*n)},\, W_{fc2} \in R^{(h*n) \times h}$.

\subsubsection{Monotonic Convolutional Multihead Attention}
We suggest the use of monotonic convolutional multihead attention. This architecture is combined with ConvBERT's \cite{jiang2020convbert} mixed-attention and AKT \cite{ghosh2020context} monotonic attention. In previous research, mixed attention achieved a higher performance than normal attention with BERT. Meanwhile, the sequence data in KT contain latent information about the forgetting of the students. To represent such forgetting, we used the exponential decay mechanism of monotonic attention. Figure \ref{monoconvattn}-\textit{Right} shows the monotonic convolutional multihead attention architecture.

The monotonic convolutional multihead attention $A_{mc}$ consists of the concatenation ($[;]$) of monotonic multihead attention $A_m$ and span-based dynamic convolution $SDC$. Here, $Am$ is the same as monotonic attention and can be formulated as follows:

\begin{equation}
    A_{mc}(Q, K, V) =[A_m(Q,K,V) ;SDC(Q, K, V)].
\end{equation}

First, monotonic multihead attention $A_m$ has an exponential decay mechanism for measuring the distance between sequences. The exponential decay mechanism is a dot product with query linear $W_Q$ and key linear $W_K$. The learnable parameter $\delta$ is multiplied by these values. In addition, $A_m$ can be formulated as follows:

\begin{equation}
    Am = softmax( \frac{(-\delta \cdot d(t, \tau)) \cdot W_Q \cdot W_K}{\sqrt{D_k}}),\, \delta > 0.
\end{equation}

Here, $d(t, \tau)$ is the distance function, where $t$ is the present time step, and $\tau$ is the previous time step. In addition, $d(t, \tau)$ can be formulated as

\begin{equation}
    d(t, \tau) = |t-\tau| \cdot \sum_{t'=\tau+1}^{t}{\gamma_{t, t'}}.
\end{equation}

Moreover, $\gamma_{t, t'}$ can be formulated as

\begin{equation}
    \gamma =  \frac{exp(\frac{W_{qt} \cdot W_{kt}'}{\sqrt{D_k}})}{\sum_{1 \leq \tau' \leq t} exp(\frac{W_{qt} \cdot W_{k\tau'}}{\sqrt{D_k}})},\, t' \leq t.
\end{equation}

The span dynamic convolution $SDC$ can be formulated as

\begin{equation}
    SDC(Q, K, V) = LConv(V, softmax(W( Q \otimes K))),
\end{equation}

where $W$ is a linear layer, and $\otimes$ can be denoted as a point-wise multiplication. The lightweight convolution $LConv$ can be formulated as follows:

\begin{equation}
    LConv(X, W) = \sum_{j=1}^{k}W_j \dot X_{i+j-[\frac{[k+1]}{2}]}
\end{equation}

\begin{table*}[hbt!]
    \renewcommand{\arraystretch}{1.5}
    \centering
    \begin{tabular}{ccccccccccc}
        \hline
        Dataset & Metrics & DKT & DKVMN & SAKT & AKT & CL4KT & MCB-NC & MCB-C \\
        \hline
        \multirow{2}{*}{assist09} & AUC & 0.7285 & 0.7271 & 0.7179 & 0.7449 & \underline{0.7600} & \textbf{0.8002} & \textbf{\underline{0.8059}}\\
                            & RMSE & \underline{0.4328} & 0.4348 & 0.4381 & 0.4413 & 0.4337 & \textbf{\underline{0.4029}} & \textbf{0.4063}\\
        \hline
        \multirow{2}{*}{assist12} & AUC & 0.7006 & 0.7011 & 0.6998 & \underline{0.7505} & 0.7314 & \textbf{0.8065} & \textbf{\underline{0.8130}}\\
                            & RMSE & 0.4348 & 0.4355 & 0.4360 & \underline{0.4250} & 0.4284 & \textbf{0.3976} & \textbf{\underline{0.3935}}\\
        \hline
        \multirow{2}{*}{assist17} & AUC & \textbf{\underline{0.7220}} & \underline{0.7095} & 0.6792 & 0.6803 & 0.6738 & 0.6700 & \textbf{0.7141}\\
                            & RMSE & \textbf{\underline{0.4469}} & \textbf{0.4516} & \underline{0.4591} & 0.4722 & 0.4713 & 0.4727 & 0.4630\\
        \hline
        \multirow{2}{*}{algebra05} & AUC & 0.8088 & 0.8146 & \underline{0.8162} & 0.7673 & 0.7871 & \textbf{0.8190} & \textbf{\underline{0.8201}}\\
                            & RMSE & 0.3703 & \underline{0.3687} & \textbf{0.3685} & 0.3918 & 0.3824 & 0.3940 & \textbf{\underline{0.3584}}\\
        \hline
        \multirow{2}{*}{algebra06} & AUC & 0.7939 & \underline{0.7961} & 0.7927 & 0.7505 & 0.7789 & \textbf{0.7997} & \textbf{\underline{0.8064}}\\
                            & RMSE & \textbf{0.3666} & \textbf{\underline{0.3661}} & 0.3675 & 0.3986 & 0.3863 & 0.3835 & \underline{0.3672}\\
        \hline
        \multirow{2}{*}{EdNet} & AUC & 0.6609 & 0.6602 & 0.6506 & \underline{0.6687} & 0.6651 & \textbf{0.7221} & \textbf{\underline{0.7336}}\\
                            & RMSE & 0.4598 & \underline{0.4597} & 0.4629 & 0.4783 & 0.4750 & \textbf{0.4572} & \textbf{\underline{0.4516}} \\
        \hline
    \end{tabular}
    \caption{Overall performance of knowledge tracing models based on five benchmark datasets. The best performance is denoted in bold underline, the second in bold, and the third in underline. MCB-C indicates that MonaCoBERT used classical test theory (CTT), whereas MCB-NC indicates that it did not. We can see that MCB-C achieved the best results, and MCB-NC was second for most of the benchmark datasets.}
    \label{tb:performance}
\end{table*}

\subsection{Experiment Setting}

\subsubsection{Datasets}
Six benchmark datasets were used to validate the effectiveness of our model. We ignored student data with fewer than five interactions. If the dataset contained multiple concepts in a single interaction, we treated the combination of concepts as unique. 
The ASSISTment datasets were collected from the ASSISTment ITS. We used assist09, assist12, assist17 and ignored assist15, which has no information regarding the questions\footnote{retrieved from https://sites.google.com/site/assistmentsdata/home}.
The algebra datasets were provided by the KDD Cup 2010 EDM Challenge\footnote{retrieved from https://pslcdatashop.web.cmu.edu/KDDCup}.
EdNet\footnote{retrieved from https://github.com/riiid/ednet} is a dataset collected by Santa for the learning of English, mainly TOEIC \cite{choi2020ednet}. We extracted 5,000 interaction data from the original dataset.
Table \ref{tb: benchmark} lists the number of features in the benchmark dataset.

\begin{table}[hbt!]
    \renewcommand{\arraystretch}{1.5}
    \captionsetup{singlelinecheck = false, justification=justified}
    \centering
    \resizebox{\columnwidth}{!}{
    \begin{tabular}{ccccc}
        \hline
        Dataset & \#Students & \#Concepts & \#Questions & \#interactions \\
        \hline
        assist09 & 3,695 & 149 & 17,728 & 282,071 \\
        assist12 & 24,429 & 264 & 51,632 & 1,968,737 \\
        assist17 & 1,708 & 411 & 3,162 & 934,638 \\
        algebra05 & 571 & 271 & 173,113 & 607,014 \\
        algebra06 & 1,318 & 1,575 & 549,821 & 1,808,533 \\
        EdNet & 5,000 & 1,472 & 11,957 & 641,712 \\
        \hline
    \end{tabular} }
    \caption{Benchmark dataset ignored student data with less than five interactions. \#Concepts are the same as the skills.}
    \label{tb: benchmark}
\end{table}

\subsubsection{Evaluation Metrics and Validation}
By referencing CL4KT, we used both the area under the curve (AUC) and the root mean squared error (RMSE) as the performance metrics. We also used a five-fold cross-validation for the evaluation.

\subsubsection{Baseline Models}
We compared MonaCoBERT to the baseline models, such as DKT \cite{piech2015dkt}, DKVMN \cite{zhang2017dynamic}, SAKT \cite{pandey2019self}, and the latest models, such as AKT \cite{ghosh2020context} and CL4KT. \cite{lee2022contrastive}.

\subsubsection{Hyperparameters for Experiments}
To compare each model, we used the same parameters for the model training.
 
\begin{itemize}
\item \textbf{batch size}: The batch size was 512. Owing to a limitation of resources, we also used a gradient accumulation.
\item \textbf{early stop}: The early stop was 10. If the validation score was not successively increased during the ten iterations, the training session was stopped.
\item \textbf{training, validation, test ratio}: The training ratio was 80\% of the entire dataset, and the test ratio was 20\%. The valid ratio was 10\% of the training ratio.
\item \textbf{learning rate and optimizer}: The learning rate was 0.001, and Adam was used as the optimizer.
\item \textbf{embedding size}: The embedding size was 512.
\item \textbf{others}: We used eight attention heads for MonaCoBERT. The Max sequence length was 100, and the encoder number was 12. Other models such as AKT\footnote{https://github.com/arghosh/AKT} and CL4KT\footnote{https://github.com/UpstageAI/cl4kt} used the default settings.
\end{itemize}

\section{Result and Discussion}

\subsection{Overall Performance}

Figure \ref{tb:performance} illustrates the overall performance of each model. Every model used a five-fold cross-validation for the estimation. MonaCoBERT-C, which was trained using CTT, was the best model in most benchmark datasets and was a new state-of-the-art model in assist09, assist12, algebra05, and ednet. MonaCoBERT-NC was the second-best model for most of the datasets. This result indicates that CTT embedding affects the performance of the model. For all datasets, MonaCoBERT-C performed better than MonaCoBERT-NC. This result indicates that it was difficult for MonaCoBERT-NC to learn the latent representations of the item difficulty from the dataset.

Our estimation differs from that of previous research. Except for MonaCoBERT-NC and MonaCoBERT-C, the best model was modified for each dataset. For instance, the AUC and RSME of assist17, and the RMSE, DKT, and DKVMN of algebra06 showed that these were the best and second-best models, respectively. This indicates that DKT and DKVMN are still helpful in predicting certain cases. These results may stem from pre-processing methods or the training of the hyperparameter settings.

\subsection{Ablation Studies}

In this section, we explore why MonaCoBERT performed better than the other models and which parts of the model affected the increase in performance.

\subsubsection{Impact of Attention Mechanisms}

In Table \ref{Comparing each attention mechanism}, we compare the attention mechanisms. For comparison, we used the assist09 and assist09-CTT datasets. The assist09 dataset is a normal dataset that contains concepts, questions, and correctness; however, assist09-CTT contains the concepts, questions, correctness, and CTT-based difficulty.

We detached each part of the monotonic convolutional multi-head attention and created four attention mechanisms: normal multi-head attention, monotonic multi-head attention, convolutional multi-head attention, and monotonic convolutional multi-head attention. We also used a five-fold cross-validation and an early stop 10 times. The other hyperparameters used to determine the overall performance were the same.

As a result, monotonic convolutional multihead attention exhibited the best performance for both comparisons. Convolutional multihead attention and monotonic multihead attention achieved the second-best performance under each setting. The increments differed for each setting and were approximately 2\% for assist09 and 1-2\% for assist09-CTT.

\begin{table}[hbt!]
    \renewcommand{\arraystretch}{1.5}
    \captionsetup{singlelinecheck = false, justification=justified}
    \centering
    \resizebox{\columnwidth}{!}{
    \begin{tabular}{ccccc}
        \hline
        Dataset & Attn & MonoAttn & ConvAttn & MonoCoAttn \\
        \hline
        assist09 & 0.7736 & \textbf{0.7993} & \underline{0.7959} & \textbf{\underline{0.8002}} \\
        increment & 0 & + 0.026 & + 0.022 & + 0.027 \\
        \hline
        assist09-CTT & 0.7858 & \underline{0.8039} & \textbf{0.8054} & \textbf{\underline{0.8059}}\\
        increment & 0 & + 0.018 & + 0.020 & + 0.021 \\
        \hline
    \end{tabular} }
    \caption{AUC performances of each attention mechanism using the assist09 and assist09-CTT datasets. The increments were written based on normal attention.}
    \label{Comparing each attention mechanism}
\end{table}

\subsubsection{Impacts of Embedding Strategy}
In Table \ref{Impacts of embedding strategy}, we compare each embedding strategy. The first embedding strategy $emb_{cq}$ is an element-wise sum of the concept embedding $emb_c$, question embedding $emb_q$, and correctness embedding $emb_r$. 

\begin{equation}\label{emb_cq}
    emb_{cq} = emb_c + emb_q + emb_r
\end{equation}

Moreover, the second embedding strategy $emb_{rasch}$ is an element-wise sum of concept and Rasch embedding, as suggested by AKT. Rasch embedding uses concept embedding $emb_c$ and learnable question scalar $emb_q$ or a combination of concepts and answer embedding $emb_{cr}$ to calculate the difficulty, where $emb_c, emb_{cr} \in R^{n \times h}$ and $emb_q \in R^{n \times 1}$. Note that IRT Rasch embedding differs from AKT Rasch embedding because the condition of IRT assumes that the knowledge state of the student is fixed and does not change when estimated.

\begin{equation}\label{rasch_c}
    emb_{rasch-c} = emb_c + emb_q * emb_c
\end{equation}

\begin{equation}\label{rasch_cr}
    emb_{rasch-cr} = emb_{cr} + emb_q * emb_{cr}
\end{equation}

The last embedding strategy, $emb_{CTT}$, is an element-wise sum of concept embedding, question embedding, correctness embedding, and CTT embedding, $emb_{ctt}$, which was suggested in this study. We set $emb_{ctt}$ as the probability of the difficulty and the integer type, where $0 \leq emb_{ctt} \leq 100$.

\begin{equation}\label{emb_ctt}
    emb_{CTT} = emb_{c} + emb_{q} + emb_{r} + emb_{ctt}
\end{equation}

As a result, in Table \ref{Impacts of embedding strategy}, $emb_{CTT}$ generally showed a better performance than the other embedding strategies. DKVMN, AKT, and MonaCoBERT performed well when using $emb_{CTT}$. This result indicates that the models did not learn the difficulty representation during training. Meanwhile, CL4KT and SAKT showed slightly better performances when using $emb_{rasch}$. DKT was not affected by the embedding strategy.

\begin{table}[hbt!]
    \renewcommand{\arraystretch}{1.5}
    \captionsetup{singlelinecheck = false, justification=justified}
    \centering
    \resizebox{\columnwidth}{!}{
    \begin{tabular}{cccc}
        \hline
        Embedding Strategy & $emb_{cq}$ & $emb_{rasch}$ & $emb_{CTT}$ \\
        \hline
        DKT & 0.7263 & \textbf{0.7274} & 0.7239 \\
        DKVMN & 0.7188 & 0.7255 & \textbf{0.7313}  \\
        SAKT & 0.6822 & \textbf{0.6941} & 0.6693  \\
        AKT & 0.7440 & 0.7449 & \textbf{0.7632} \\
        CL4KT & 0.7600 & \textbf{0.7601} & 0.7461  \\
        MCB & 0.8002 & 0.7736 & \textbf{0.8059} \\
        \hline
    \end{tabular} }
    \caption{Comparison of each embedding strategy with KT models in the assist09 dataset.}
    \label{Impacts of embedding strategy}
\end{table}

\subsection{In-depth Analysis of Attention and Embedding}

In this subsection, we analyze the attention and embedding in depth. We used Grad-CAM and t-SNE for the analysis and visualization.

\begin{figure*}[t]
    \centering
        \begin{subfigure}
            \centering
            \includegraphics[width=0.3\textwidth]{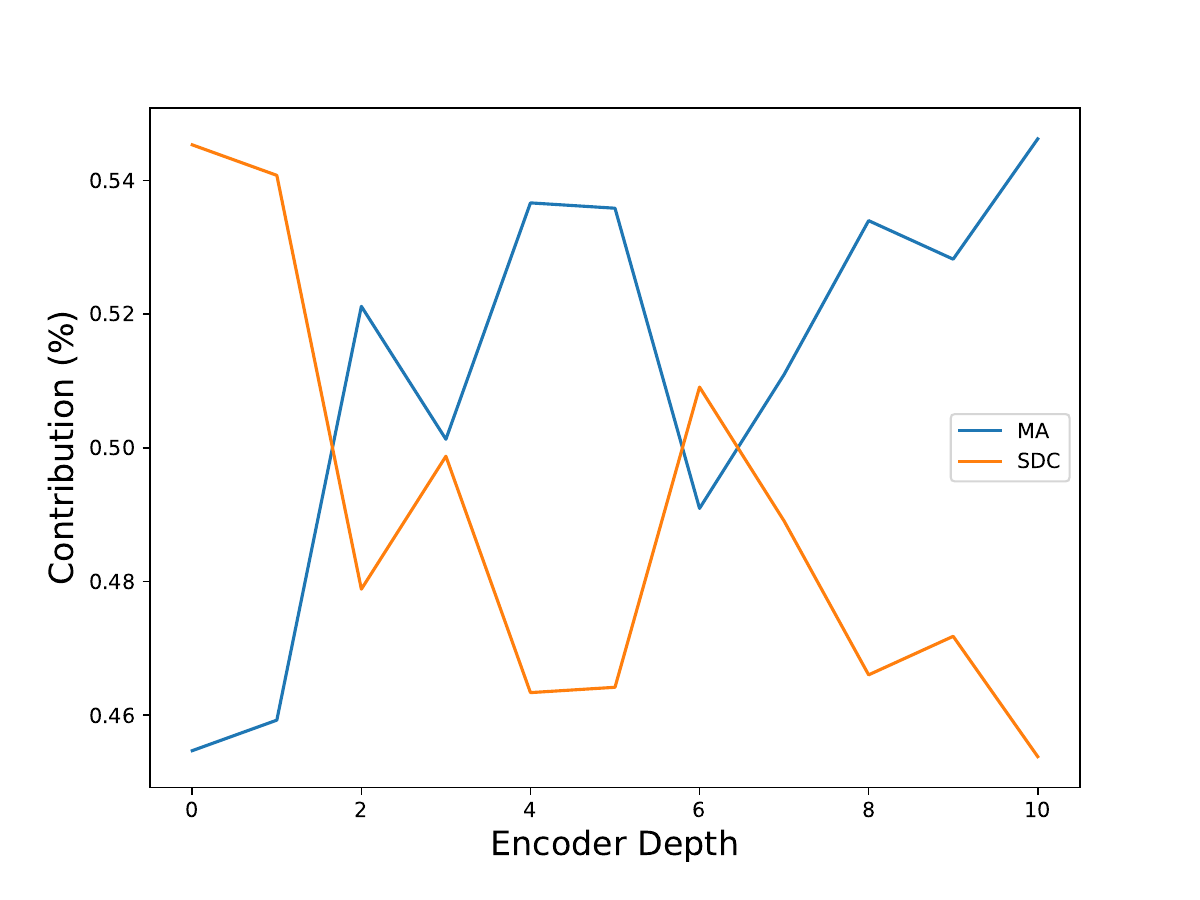}
        \end{subfigure}
        \begin{subfigure}
            \centering
            \includegraphics[width=0.3\textwidth]{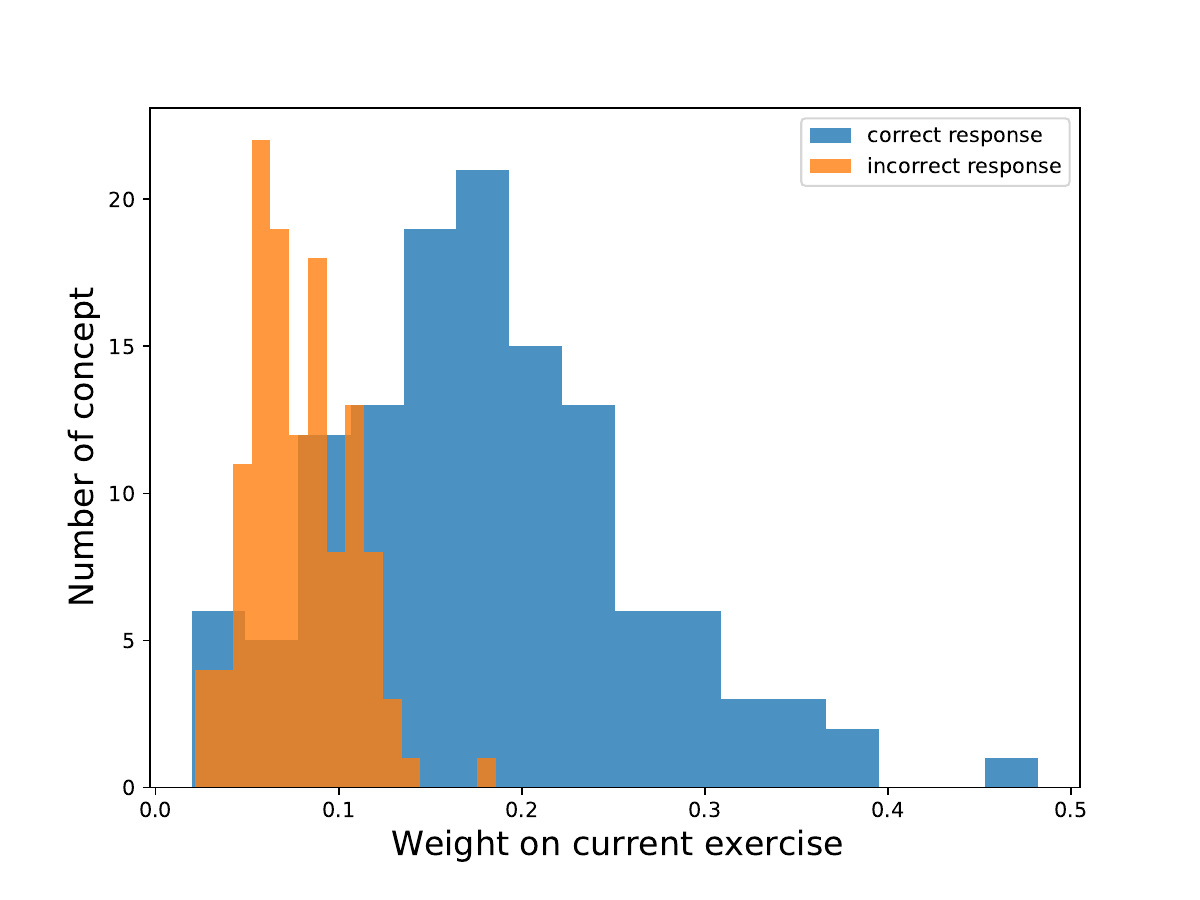}
        \end{subfigure}
        \begin{subfigure}
            \centering
            \includegraphics[width=0.3\textwidth]{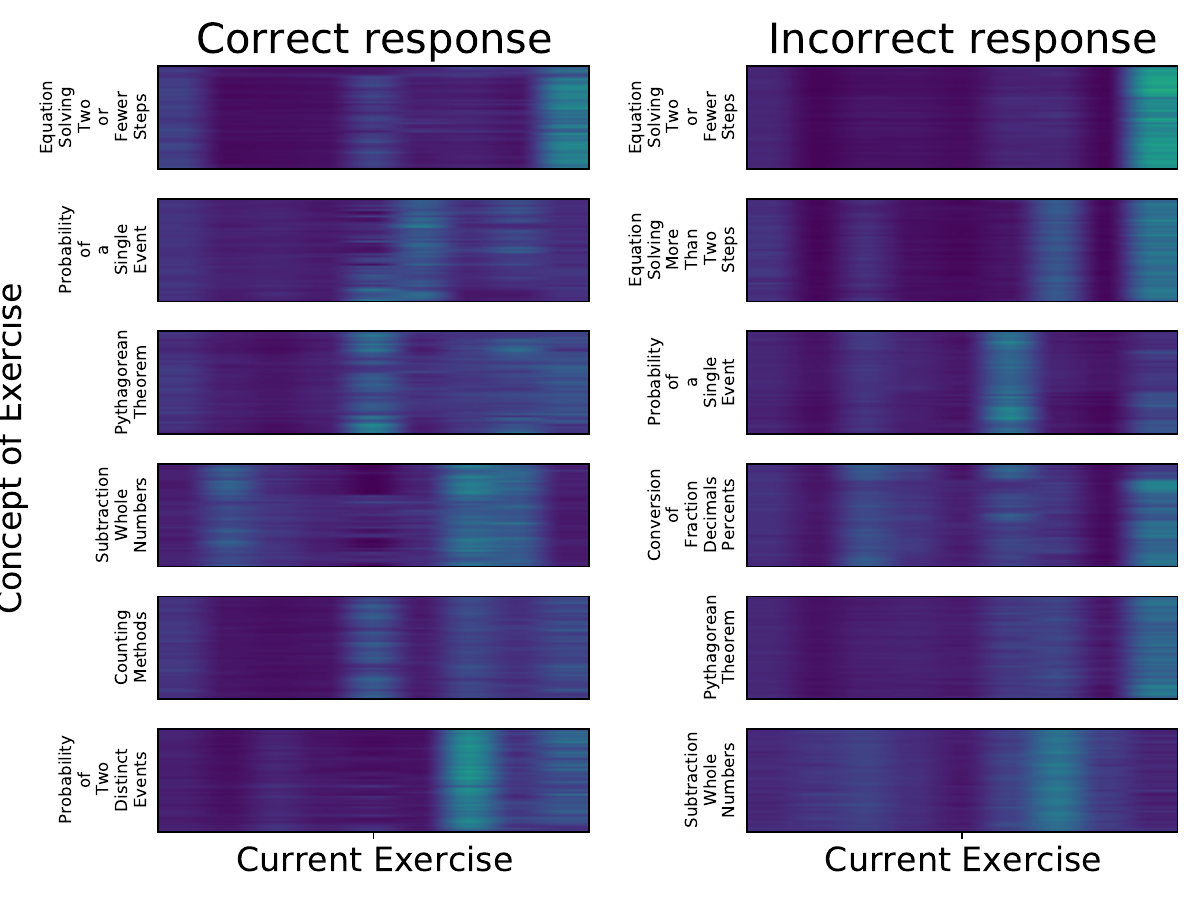}
        \end{subfigure}
    \caption{Analysis of underlying behaviors of SDC. The figure on the \textit{left} illustrates the proportion of the importance of each module. SDC showed an importance competitive to that of MA in most layers. In particular, the SDC showed the most significant contribution in the first layer. The histogram in the \textit{center} figure represents the current input weight of the concept. When the response of the student was correct, the SDC allocated more weight to the interaction. In addition, even if the response was the same, the weight varied considerably based on the concept. The figure on the \textit{right} shows examples of SDC filters arranged based on the correctness and concept.} 
    \label{SDC_XAI}
\end{figure*}


\subsubsection{Analysis of MA and SDC}
\begin{figure}[hbt!]
    \centering
        \begin{subfigure}
            \centering
            \includegraphics[width=0.2\textwidth]{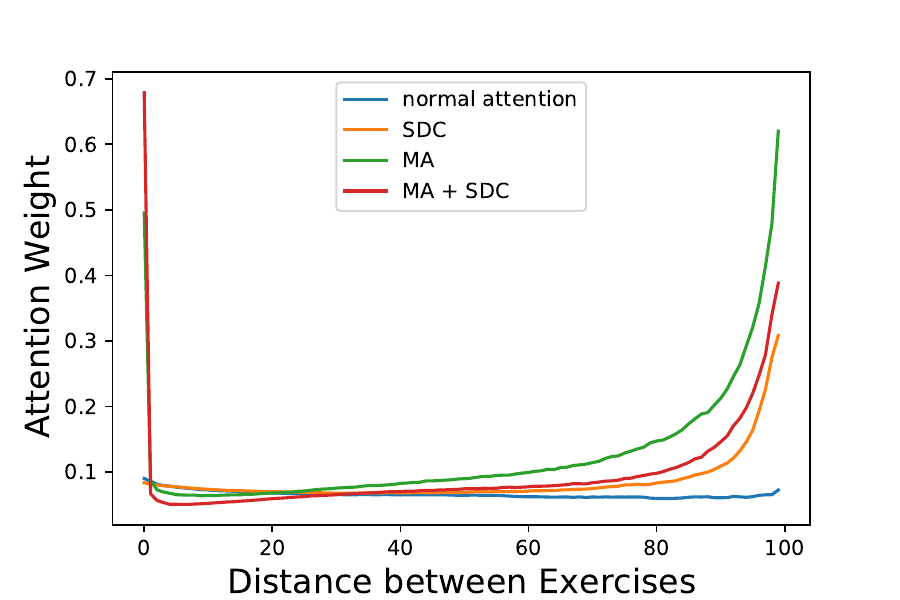}
        \end{subfigure}%
        \begin{subfigure}
            \centering
            \includegraphics[width=0.2\textwidth]{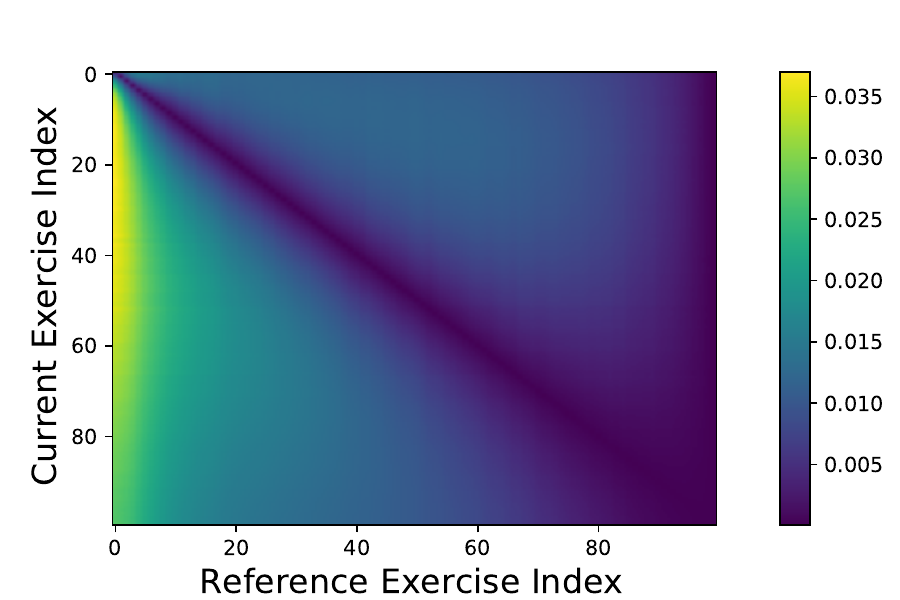}
        \end{subfigure}%
    \caption{Analysis of the attention map of monotonic self-attention (MA). The figure on the \textit{left} shows the attention weight according to the distance of the interaction data, and indicates that models with MA (e.g., SDC + MA, MA) display more outstanding attention scores for distant tokens. The figure on the \textit{right} shows an example of the attention map of MonaCoBERT.}
    \label{MA_attn_map}
\end{figure}

Owing to the nature of KT, such as the forgetting behavior of the students, we expected that monotonic attention (MA) will look up the nearby data regarding the current input. However, as shown in Figure \ref{MA_attn_map}, MA induced higher attention scores for the distant data, not nearby data. We also observed that SDC was more critical than MA in the first layer. Figure \ref{SDC_XAI}-\textit{Left} shows the relative importance ratios of SDC and MA. The contribution of SDC was greater than that of MA in the first layer. To define the importance of each module, we used an element-wise version of Grad-CAM as a metric \cite{selvaraju2017grad, jacobgilpytorchcam}. 
We also found that SDC extracted useful information regarding the properties of the current input. Specifically, SDC focused on the current input when the student answered correctly. In Figure \ref{SDC_XAI}-\textit{Center, Right}, we can see that SDC assigned higher weights to the current inputs when the student responded correctly. Moreover, the large variance of weights given correct responses implies that SDC considers not only the correctness of responses but also the importance of the concept. (Figure \ref{SDC_XAI}-\textit{Center}, blue) This result shows us that MonaCoBERT implicitly learned what concepts or questions were essential for inferring the ability of the students. This indicates the possibility of using MonaCoBERT to automatically find the problem essential to estimating the student's ability, which can be used to support the estimation and assessment.

\subsubsection{CTT based embedding}

\begin{figure}[hbt!]
    \centering
    \includegraphics[width=0.5\textwidth]{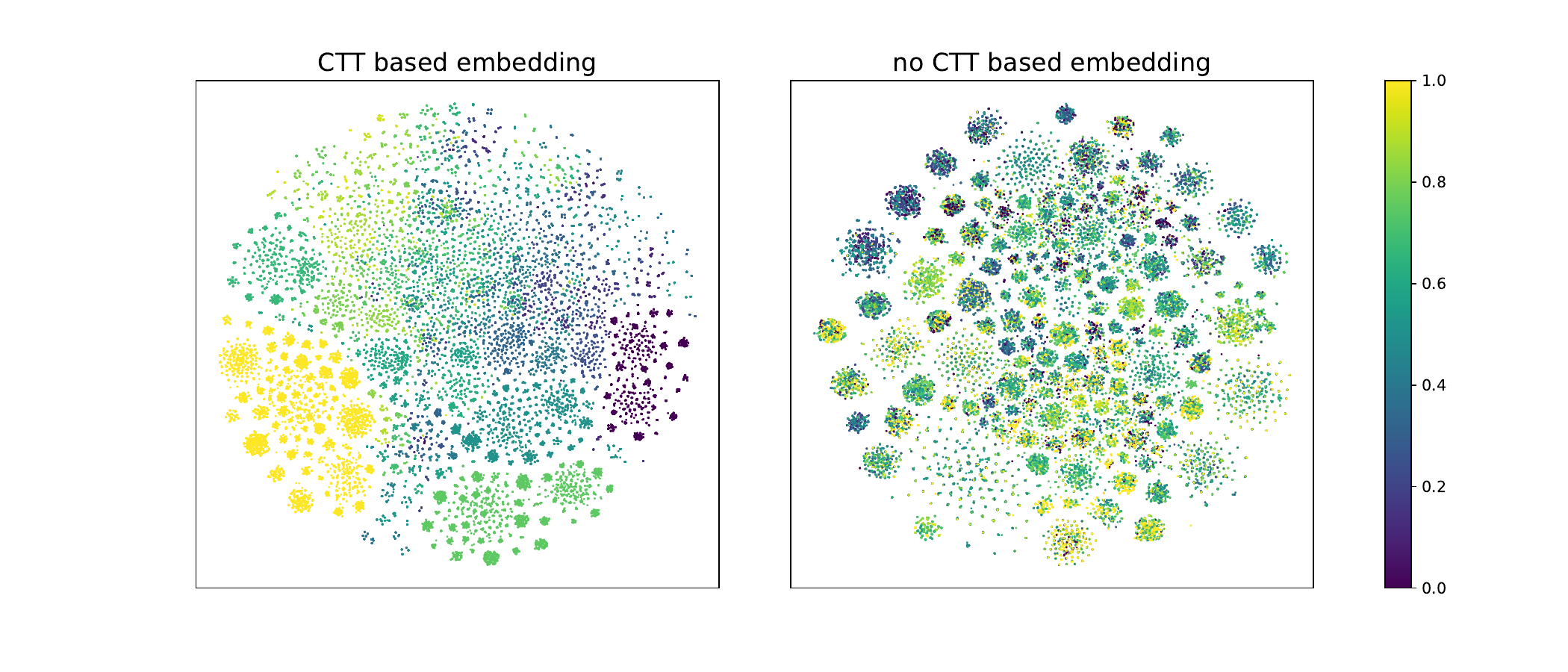}
    \caption{Visualization of the embedding vector. The figure on the \textit{left} shows the results with CTT-based embedding. The figure on the \textit{right} shows the results of No-CTT-based embedding. We can see that the results of CTT-based embedding not only represent the difficulty information globally, they also help avoid a difficulty in the mixing in each cluster.}
    \label{emb_vis}
\end{figure}

We showed that CTT-based embedding helps the model represent the difficulty of the problem. Figure \ref{emb_vis} shows a visualization using t-SNE \cite{van2008visualizing}. Figure \ref{emb_vis}-\textit{Left} shows the visualization of the CTT-based embedding vector, and Figure \ref{emb_vis}-\textit{Right} shows the visualization of the No-CTT-based embedding. Unlike No-CTT-based embedding, where different difficulties are mixed in each cluster, CTT-based embedding (i.e., $emb_{CTT}$) showed that the difficulty of the information was smoothly distributed globally.

\subsection{Discovery of Relationships between Concepts}

\begin{figure*}[t]
    \centering
        \begin{subfigure}
            \centering
            \includegraphics[width=0.35\textwidth]{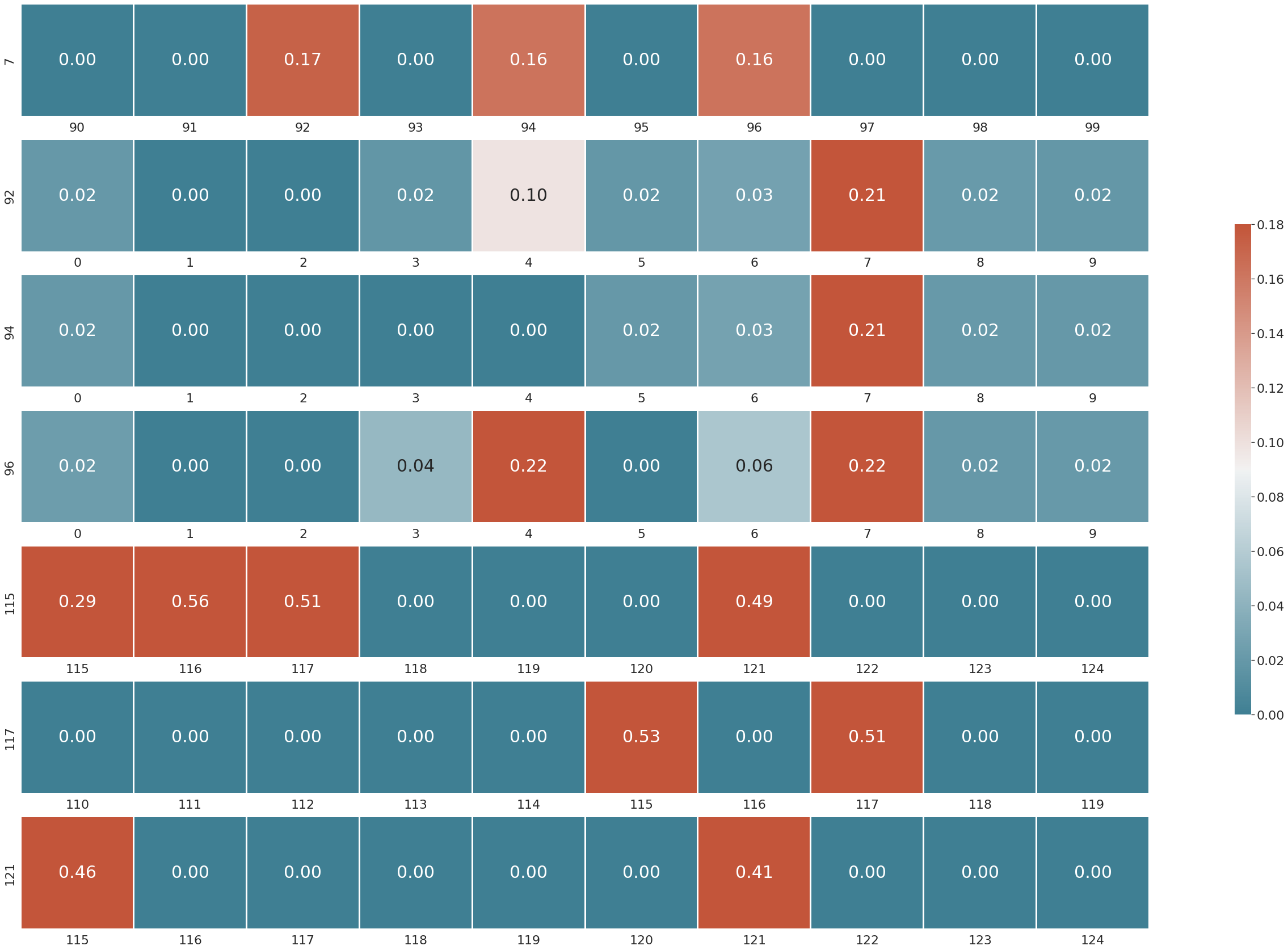}
        \end{subfigure}
        \begin{subfigure}
            \centering
            \includegraphics[width=0.3\textwidth]{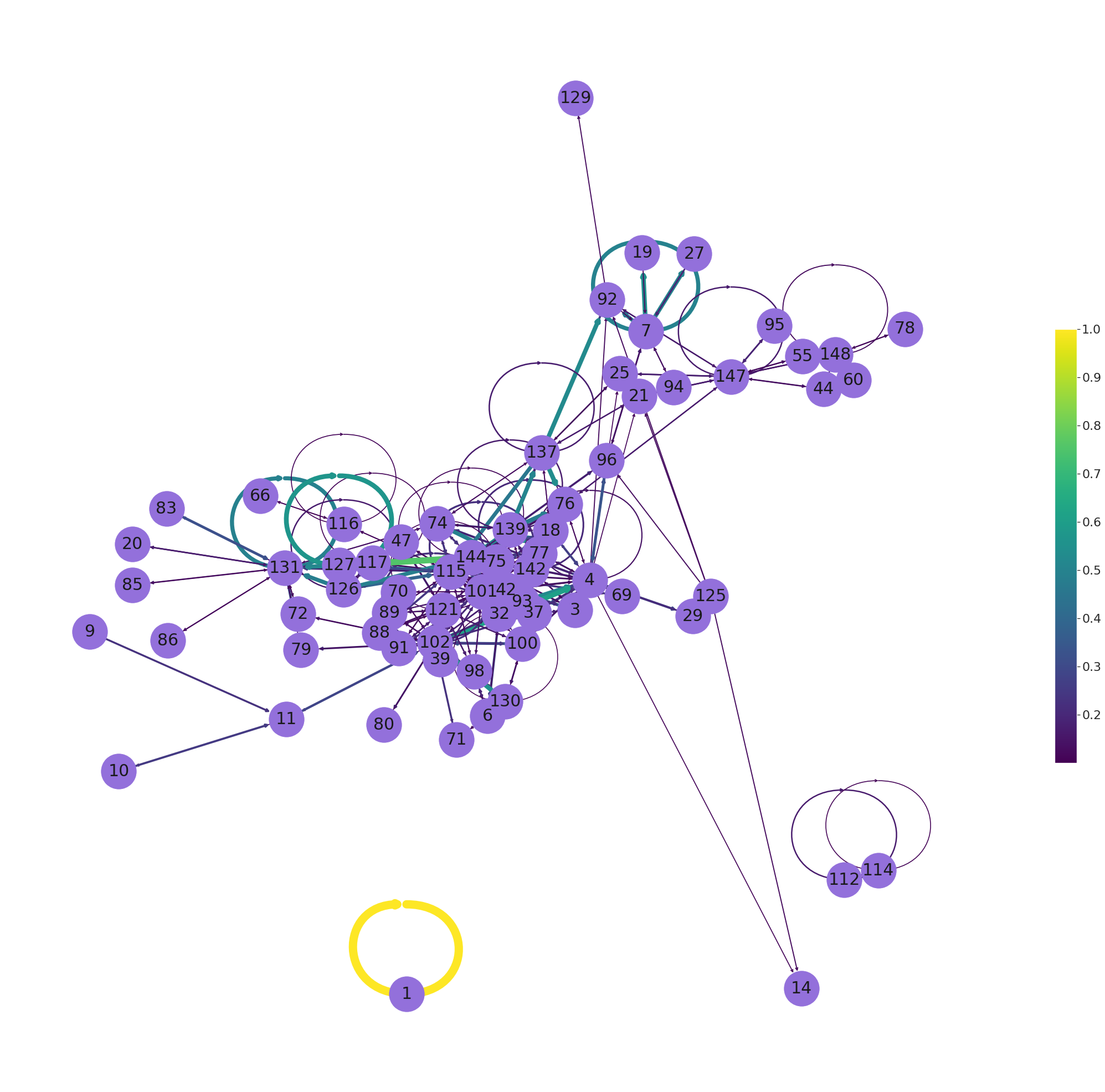}
        \end{subfigure}
        \begin{subfigure}
            \centering
            \includegraphics[width=0.3\textwidth]{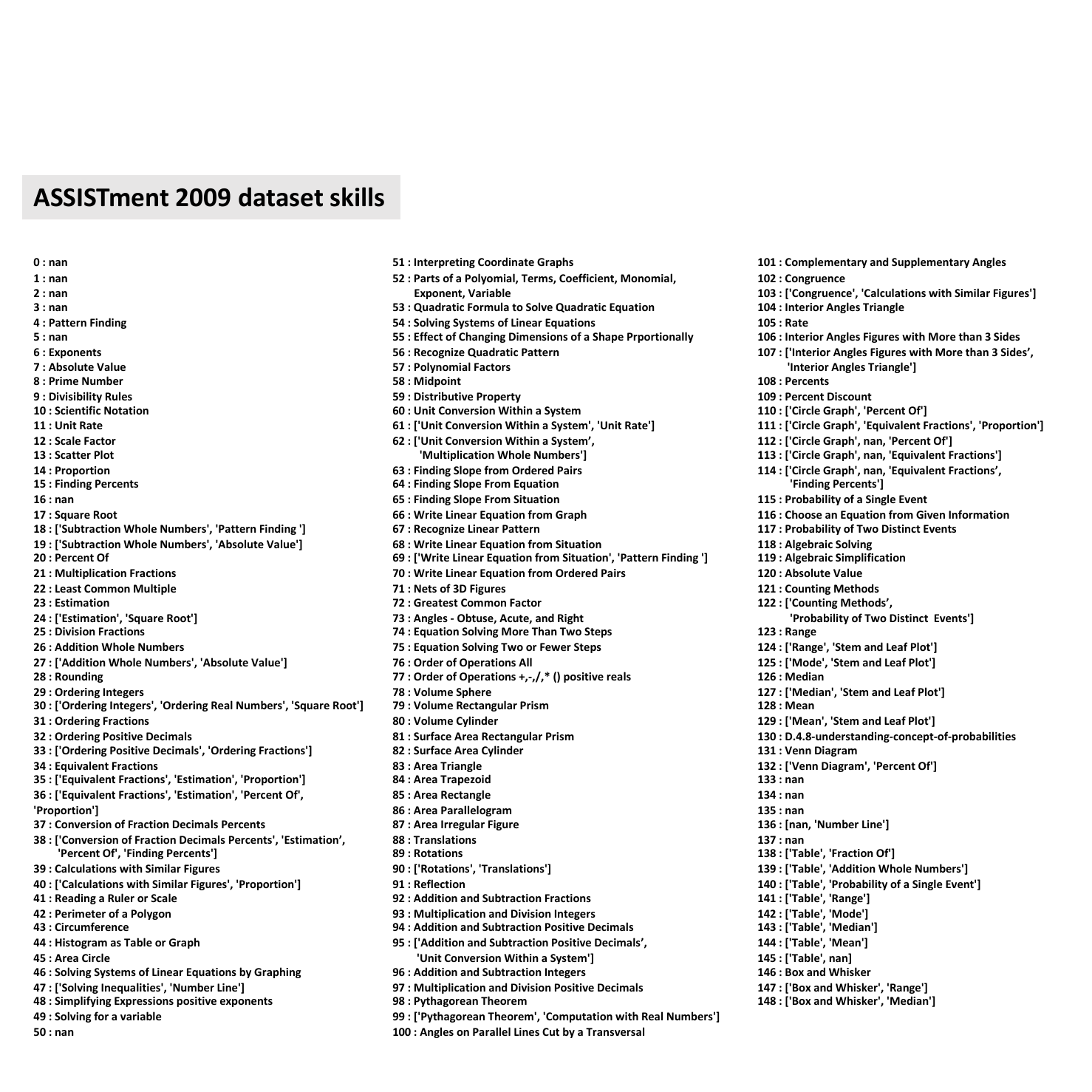}
        \end{subfigure}
    \caption{
    Analysis results of the relevance between concepts, exploiting attention weights of the monotonic attention part after the model was trained using monotonic convolutional multi-head attention.
    The figure on the \textit{left} shows a heatmap of the attention weights between each pair of concepts. It shows how much attention each concept on the y-axis (e.g., 7th, 92nd, 94th, 96th) assigns attention to some selected concept on the other x-axis.
    The \textit{center} figure shows a directed graph of the relevance between concepts. It shows how the concepts of assist09 influence one another. The source concept nodes are assigned a high attention weight to the destination concept nodes, and the concept nodes can be connected in both directions. We set the threshold to 0.1 and ignored edges lower than the threshold. When the threshold was decreased, more skill nodes were connected, and vice versa.
     The concept information of the assist09 dataset can be found on the \textit{right}. `nan' means concepts that are not defined in the original dataset.
     } 
    \label{attnvis}
\end{figure*}

To determine whether our model understood the relevance between concepts, we analyzed the monotonic attention weights of the last encoder layer after passing through the softmax function. The results are shown in Figure \ref{attnvis}-\textit{left}. We averaged the attention scores of the questions using the same concepts to obtain the relevance between concepts. We created a directed graph, as shown in Fig \ref{attnvis}-\textit{Center}, by selecting only those concepts with attention weights of higher than 0.1.

According to the concept network shown in Figure \ref{attnvis}-\textit{center}, we can see that the model learns the relevance between skills. For example, as shown in Figure \ref{attnvis}-\textit{left}, the 7th concept (Absolute Value) 
was connected with some concepts of subtraction, such as 92 (Addition and Subtraction Fractions), 94 (Addition and Subtraction Positive Decimals), and 96 (Addition and Subtraction Integers). This means that you need to be good at subtraction to calculate the correct absolute value. Accordingly, the 117th concept (Probability of a Single Event) and 115th concept (Probability of Two Distinct Events) assigned high attention weights to each other, since concept 117 is a prerequisite for concept 115. 121st concept (Counting Methods) is also connected with 115 and 117. 
However, the concept network shown in Figure \ref{attnvis} is not perfect because some concepts did not connect to each other despite their similarities. This result may be due to the monotonic attention decreasing the attention weight according to the time step. Nevertheless, observing the attention weights can help uncover new connections between previously inconceivable concepts.

\section{Conclusion}

In this study, we developed MonaCoBERT, which employs a BERT-based architecture with monotonic convolutional multihead attention for student forgetting and the representation power of the model. We also adopted an effective embedding strategy that represented difficulty based on a classic test theory. Consequently, MonaCoBERT exhibited superior performance on most benchmark datasets. We conducted an ablation study for each part of the model; consequently, we discovered that monotonic convolutional multihead attention aided in improving the model performance. Although the embedding strategy contributed significantly to the performance improvement of our model, we confirmed that depending on the model, the contribution of the embedding strategy to the performance enhancement differed. We conducted an additional analysis to quantitatively analyze the attention architecture and embedding strategy using Grad-CAM and t-SNE. Future research will be focused on improving the attention architecture and the difficulty embedding strategy.

\bibliography{0_main.bbl}

\end{document}